\documentclass[12pt]{article}
\usepackage{latexsym}
\usepackage{amsmath}
\usepackage{amssymb}
\usepackage{amsthm}
\usepackage{graphicx}
\usepackage{hyperref}
\usepackage[latin1]{inputenc}
\usepackage[T1]{fontenc}
\usepackage[all]{xy}
\usepackage{array,multirow}
\makeatletter
\renewcommand{\@seccntformat}[1]
{\csname the#1\endcsname.\enspace} \makeatother
\def \XXint#1#2#3{{\setbox0=\hbox{$#1{#2#3}{\int}$}
     \vcenter{\hbox{$#2#3$}}\kern-.5\wd0}}

\newtheorem{theorem}{Theorem}
\newtheorem{lemma}{Lemma}

\newtheorem{corollary}{Corollary}
\newtheorem{example}{Example}
\newtheorem{definition}{Definition}

\begin{document}
\begin{center}
   {\bf An Optimal Combination of Proportional and Stop-Loss Reinsurance Contracts From Insurer's and Reinsurer's Viewpoints \footnote{\today}}\\
{\sc {Amir T. Payandeh-Najafabadi}}\footnote{Corresponding author:
amirtpayandeh@sbu.ac.ir; Phone No. +98-21-29903011; Fax No. +98-21-22431649}\& Ali Panahi-Bazaz\\
Department of Mathematical Sciences, Shahid Beheshti University,
 Evin, Tehran, Iran, 19839-63113.
\end{center}
\begin{center}
    {\sc Abstract}
\end{center}
A reinsurance contract should address the conflicting interests of
the insurer and reinsurer. Most of existing optimal reinsurance
contracts only considers the interests of one party. This article
combines the proportional and stop-loss reinsurance contracts and
introduces a new reinsurance contract called
proportional-stop-loss reinsurance. Using the balanced loss
function, unknown parameters of the proportional-stop-loss
reinsurance have been estimated such that the expected surplus for
both the insurer and reinsurer are maximized. Several
characteristics for the new reinsurance are provided.\\
\textbf{\emph{Keywords:}} Proportional reinsurance; Stop-loss
reinsurance; Expected utility; Bayesian approach; Balanced loss
function.
\section{Introduction}

Designing an optimal reinsurance strategy is an interesting
actuarial problem that must balance several conflicting interests.
Most of existing optimal reinsurance strategies only considers the
interest of one side. Gerber (1979) showed that excess of loss
reinsurance maximizes the adjustment coefficient when the loading
coefficient is independent of the type of reinsurance strategy and
the reinsurance premium calculation principle used is the expected
value principle.

Other authors have reached similar results for reinsurance that
favor the insurance company. Khan (1961), Arrow (1963; 1974),
Beard et al. (1977), Cai \& Tan (2007), Cai et al. (2008) and Tan
et al. (2011) all represent the perspective of the insurance
company. All research represents one side of the reinsurance
contract, but because of the nature of reinsurance contracts, both
the insurance and reinsurance companies must be represented. Borch
(1960) discussed optimal quota-share retention and stop-loss
retention to maximize the product of the expected utility
functions of two-party profits. Similar results in favor of two
parties were developed by Borch (1969), Ignatov et al. (2004),
Kaishev \& Dimitrova (2006), Dimitrova \& Kaishev (2010), and Cai
et al. (2013).

Some researchers have achieved a balance between desirability of
the insurance and reinsurance companies by combining different
reinsurance strategies. This approach began with Centeno (1985),
who combined quota-share and excess of loss reinsurance strategies
and defined a new reinsurance strategy. She assumed that the
insurance company will pay $\min \{ \alpha X,M\} $ for loss $X$
and constant $\alpha $ and $M$. She estimated $\alpha $ and $M$ by
minimizing the coefficient of variation and the skewness of the
insurance loss. Centeno \& Simo\~es (1991) determined parameters
for a mixture of quota-share and excess of loss reinsurance so
that adjustment coefficient $R$ is maximized. Liang and Guo (2011)
used the reinsurance strategy proposed by Centeno (1985) and
estimated $\alpha $ and $M$ by maximizing the expected exponential
utility from terminal wealth.

Gajek \& Zagrodny (2000) showed that for a bounded-above
reinsurance premium, the reinsurance strategy that minimizes the
variance of the retained risk of the insurance company takes the
form $(1-\alpha )(X-M)I_{[M,\infty )} (X)$ as the reinsurance
portion of loss $X$. Kaluszka (2004) derived an optimal
reinsurance strategy that is a trade-off for the insurer between
decreasing the variance of the retained risk and the expected
value of its gain. Guerra \& Centeno (2008) provided optimal
reinsurance that maximizes the adjustment coefficient of the
retained risk by exploring the relationship between the adjustment
coefficient and expected wealth exponential utility. Cai et al.
(2013) and Fang \& Qu (2014) examined the reinsurance strategy of
Centeno (1985). They maximized the joint survival probability of
both the insurer and reinsurer and derived a class of estimators
for the parameters of the reinsurance strategy.

These results and those of other studies may lead one to conclude
that optimality for a reinsurance strategy is either finding a
strategy between all possible (or constrained) reinsurance
strategies or estimating unknown parameters of a given reinsurance
strategy. The present article defines optimal reinsurance by
estimating unknown parameters $\alpha $ and $M$ as:
\begin{eqnarray}
\label{Model-New-reinsurance} {Y_i} = \alpha \min \left( {{X_i},M}
\right)
\end{eqnarray}
which is the insurer portion from random claim $X_{i} $ under a
reinsurance strategy. This form of reinsurance strategy is called
proportional excess of loss reinsurance and is a version of the
reinsurance from Centeno (1985). More precisely, she considered
$Y_i = \min \{\alpha X_i,M \}$ as the insurer portion from random
claim $X_{i}.$ Therefore, one may conclude that, there is not any
essential difference between reinsurance strategy
\eqref{Model-New-reinsurance} and Centeno (1985). But, this
article estimates two unknown parameters the new strategy
\eqref{Model-New-reinsurance} by taking into account both parties
(i.e., insurer's and reinsurer's companies). More precisely,
unknown parameters $\alpha $ and $M$ in the proportional excess of
loss reinsurance strategy shown in Equation
\eqref{Model-New-reinsurance} can be estimated in two steps.
First, estimate the parameters such that the expected utility of
the insurer (or reinsurer) is maximized. Next, use the estimated
parameters from the insurer and reinsurer as target estimators.
Then develop a Bayesian estimator with respect to the
doubly-balanced loss function for each parameter so that the
expected surplus of the insurer and reinsurer are maximized.

Section 2 defines elements of the proposed method. Section 3
examines optimal properties of the proportional excess of loss
reinsurance strategy. The Bayesian estimator for a doubly-balanced
loss function for the parameters of the proportional-excess-loss
reinsurance strategy are described in Section 4. Section 5
provides an example of practical implementation of the results.
Section 6 concludes the paper.
\section{Preliminaries and Model}
Suppose random claim $X_{i} $ has cumulative distribution function
$F(x),$ and survival function $\bar{F}(x).$ Moreover, suppose that
random claim $X_{i} $ can be decomposed to the sum of the insurer
portion ($Y_{i} $) and reinsurer portion ($I (X_{i} )$), i.e.,
$X_{i} =Y_{i} +I(X_{i} )$. Now consider the combination of the
proportional and excess of loss reinsurance strategies, such as
proportional-excess-loss reinsurance $Y_{i} =\alpha \min
\left(X_{i} ,M\right).$

Next define the value-at-risk (VaR) and tail-value-at-risk (TVaR), the most popular risk measures.

\begin{definition}
\label{VaR} Suppose $X$ stands for a random risk. The
Value-at-Risk and the Tail-Value-at-Risk at level $p \in (0,1)$,
are defined as:
\begin{eqnarray*}
  \hbox{VaR}[X;p] &=& \inf \{ x \in R\left| {{F_X}(x) \ge p}
\right.\}  ; \\
  \hbox{TVaR}[X;p] &=& \frac{1}{{1 - p}}\int_p^1 {\hbox{VaR}[X;\xi ]d\xi
  },
\end{eqnarray*}
where $F(x)$ stands for the cumulative distribution function of
$X.$
\end{definition}
Random variable $X$ is less dangerous than random variable
\textit{Y} whenever ${\kern 1pt} VaR{\kern 1pt} [X;\alpha _{0}
]\le {\kern 1pt} VaR{\kern 1pt} [Y;\alpha _{0} ]$ for given
probability level $\alpha _{0} \in (0,1).$ TVaR is the arithmetic
average of the VaRs of $X$ from $p$ to $1.$ The VaR at given level
$p$ does not provide useful information about the thickness of
$X$, but TVaR does (Denuit et al. 2005).
The following represents definition of the ordinary balanced loss
function for given target estimators $\delta _{0} $ and $\delta
_{1} ,$ a doubly-balanced loss function. The target estimator is a
well-known value for a specific parameter.
\begin{definition}
\label{balanced-loss-function} Suppose $\delta _0$ and $\delta_1$
are given target estimators for unknown parameter $\xi.$
Moreover, suppose that $\rho(\cdot,\cdot)$ is an arbitrary and
given loss function. The doubly-balanced loss function of the
measure of closeness of estimator $\delta$ to target estimators
$\delta _0$ and $\delta_1$ and unknown parameter $\xi$ under loss
function $\rho(\cdot,\cdot)$ is
\begin{eqnarray}
\label{doubly-balance-loss}
\nonumber
{L_{\rho ,{\omega _1},{\omega
_2},{\delta _0},{\delta _1}}}(\xi ,\delta ) = &&{\omega _1}\rho
({\delta _0},\delta ) + {\omega _2}\rho ({\delta _1},\delta ) \\
&&+ (1
- {\omega _1} - {\omega _2})\rho (\xi ,\delta ),
\end{eqnarray}
where $\omega_1\in [0,1)$ and $\omega_2\in[0,1)$ are weights which satisfy $\omega_1 + \omega_2 < 1.$
\end{definition}

The ordinary balanced loss function with one given target
estimator was introduced by Zellner (1994) and improved by Jafari
et al. (2006), among others. For convenience, $L_{0} $ will
subsequently be used instead of $L_{\rho ,0,0,\delta _{0} ,\delta
_{1} } $ whenever $\omega _{1} =0$ and $\omega _{2} =0.$ Theorem
\eqref{double balance} derives a Bayesian estimator for $\xi $
under the doubly-balanced loss function $L_{\rho ,\omega _{1}
,\omega _{2} ,\delta _{0} ,\delta _{1} }.$
\begin{theorem}
\label{double balance} Suppose expected posterior losses $\rho
({\delta _0},\delta )$ and $\rho ({\delta _1},\delta )$ are
finite for at least one $\delta$ in which $\delta \ne {\delta _i},$
for $i=0,1.$ The Bayesian estimator for $\xi$ for prior distribution $\pi (\xi)$ and under $L_{\rho ,{\omega _1},{\omega
_2},{\delta _0},{\delta _1}}$ is equivalent to the Bayesian estimator for prior distribution:
\begin{eqnarray*}
{\pi ^*}(\xi \left| x \right.) = &&{\omega _1}{1_{\{ {\delta_0}(x)\} }{{(\xi)}}} + {\omega _2}{1_{\{ {\delta _1}(x)\} }}{{(\xi)}} \\
&&+(1 - {\omega
_1} - {\omega _2})\pi (\xi \left| x \right.),
\end{eqnarray*}
under loss function $L_0:=L_{\rho ,{0},{0},{\delta _0},{\delta
_1}}.$
\end{theorem}
\textbf{Proof.} Suppose that measures ${\mu _X}( \cdot )$ and
${\mu' _X}( \cdot )$ dominate ${\pi }(\xi \left| x \right.)$ and
${\pi ^*}(\xi \left| x \right.),$ respectively. By the definition of
Bayesian estimators under finite expected posterior loss $\rho
({\delta _0},\delta )$ and $\rho ({\delta _1},\delta ):$
\begin{eqnarray*}
  \arg {\min _\delta }&&\int_\Xi  {\{ {\omega _1}\rho ({\delta _0},\delta ) + {\omega _2}\rho ({\delta _1},\delta )}\\
&&~~{ + (1 - {\omega _1} - {\omega _2})\rho (\xi ,\delta )\} \pi (\xi \left| x \right.)d{\mu _X}(\xi )} \\
= \arg {\min _\delta }&&\int_\Xi  {\{ {\omega _1}\rho (\xi ,\delta ){1_{\{ {\delta _0}(x)\} }}(\xi) + {\omega _2}\rho (\xi ,\delta ){1_{\{ {\delta _1}(x)\} }}(\xi)}\\
    &&~~ {+ (1 - {\omega _1} - {\omega _2})\rho (\xi ,\delta )\} \pi (\xi \left| x \right.)d{\mu _X}(\xi )} \\
= \arg {\min _\delta }&&\int_\Xi  {\rho (\xi ,\delta )\{ {\omega _1}{1_{\{ {\delta _0}(x)\} }}(\xi) + {\omega _2}{1_{\{ {\delta _1}(x)\} }}(\xi)}\\
&&~~{ + (1 - {\omega _1} - {\omega _2})\} \pi (\xi \left| x \right.)d{\mu _X}(\xi )} \\
= \arg {\min _\delta }&&\int_{\Xi  \cup \{ {\delta _0}(x)\}  \cup
\{ {\delta _1}(x)\} } {{L_0}(\xi ,\delta ){\pi ^*}(\xi \left| x
\right.)d{{\mu '}_X}(\xi )}\\
=\delta^* (x). ~~&&\square
\end{eqnarray*}
This theorem is an extension of Lemma (1) in Jafari et al. (2006).
The next corollary provides a Bayesian estimator under the
doubly-balanced loss function with square error loss.
\begin{corollary}
\label{square loss} The Bayesian estimator for prior $\pi$
and under the doubly-balanced loss function with square error loss
($\rho (\xi ,\delta ) = {(\xi - \delta )^2}$) is the square
error doubly-balanced loss function given by:
\begin{eqnarray}
\nonumber
\label{double square} {\delta _{\pi ,{\omega _1},{\omega _2}}}(x)
= {E_{{\pi ^*}}}(\xi \left| x \right.) &=&{\omega _1}{\delta _0}(x)
+ {\omega _2}{\delta _1}(x)\\
&& + (1 - {\omega _1} - {\omega
_2}){E_\pi }(\xi \left| x \right.).
\end{eqnarray}
\end{corollary}
\section{Optimal properties of proportional-excess-loss reinsurance}
This section considers the proportional-excess-loss reinsurance in
Equation \eqref{Model-New-reinsurance} and establishes appropriate
properties for that reinsurance strategy. Theorem \eqref{optim 1}
shows that the proportional-excess-loss reinsurance minimizes the
variance of the retained risk in some situations.
\begin{theorem}
\label{optim 1} Suppose $I(X)$ and $I_N (X)$ are the reinsurer
contribution under an arbitrary reinsurance strategy and the
proportional-excess-loss reinsurance for random claim $X,$
respectively. Moreover, suppose that $E(I(X))= E(I_N (X))$ and
\begin{description}
    \item[(i)]   $P(I(X) \ge I_N (X)|X \le M)=1;$
    \item[(ii)] $P(I(X) \ge I_N (X)|X \ge M \& X-I(X) \le M)=1;$
    \item[(iii)] $P(I(X) \le I_N (X)|X \ge M \& X-I(X) \ge M)=1;$
\end{description}
Then variance of the retained risk under the
proportional-excess-loss reinsurance is less than such arbitrary
reinsurance strategy, i.e., $Var(X - I(X)) \ge Var(X - {I_N}(X)).$
\end{theorem}
\textbf{Proof.} When $E(I(X)) = E(I_N (X))$, $Var(X - I(X)) \ge
Var(X - {I_N}(X))$ whenever $E[(X - I{(X)})^2] \ge E[(X -
{I_N}{(X)})^2]$. Setting $W(X):= X - {I_N}(X) - M$ and $V(X):= X -
I(X) - M.$ Since $E(W(X))=E(V(X))$, it suffices to show that
$\left| {V(X)} \right| \ge \left| {W(X)} \right|$ with probability
one. Now consider the following cases:
\begin{description}
    \item[(i)] If $X \le M$ then $W(X)<0,$
 \begin{eqnarray*}
  \left| {V(X)} \right| \ge \left| {W(X)} \right| &\Leftrightarrow& \left| {X - I(X) - M} \right| \ge M - \alpha X
  \\
  &\Leftrightarrow& M + I(X) - X \ge M - \alpha X\\
  & \Leftrightarrow& (1 - \alpha )X \le I(X)\\
  &\Leftrightarrow& {I_N}(X) \le I(X);
\end{eqnarray*}
    \item[(ii)] If $X > M$ then $W(X)<0,$
\begin{eqnarray*}
  \left| {V(X)} \right| \ge \left| {W(X)} \right|}{
  \\ &\Leftrightarrow& \left| {X - I(X) - M} \right| \ge (1 - \alpha
  )M\\
  &\Leftrightarrow& \left\{ \begin{gathered}
  M - X + I(X) \ge (1 -\alpha )M~\hbox{for~}X - I(X) < M; \hfill \\
 X - I(X) - M \ge (1 - \alpha )M~\hbox{for~}X - I(X) > M, \hfill \\
\end{gathered}  \right.\\
&\Leftrightarrow& \left\{ \begin{gathered}
  I(X) \ge {I_N}(X),~~~~~~~~~~~~~~~~~~~\hbox{for~}X - I(X) < M; \hfill \\
 {I_N}(X) \ge I(X),~~~~~~~~~~~~~~~~~~~\hbox{for~}X - I(X) > M, \hfill \\
\end{gathered}  \right..~\square
\end{eqnarray*}
\end{description}
Theorem \eqref{optim 1} provides conditions under which variance
of the insurer contribution under proportional-excess-loss
reinsurance is less than under other reinsurance strategies.
Excess of loss and proportional reinsurance strategies do not
satisfy Theorem \eqref{optim 1} conditions. Therefore, the above
finding does not contradict with Bowers et al. (1997).

The following theorem compares proportional-excess-loss
reinsurance with the proportional reinsurance and the excess of
loss reinsurance strategies for stochastic dominance.
\begin{theorem}
\label{compare-Stop-Proportiona-new} Suppose $I_N(X)$ is
the contribution of reinsurance against random claim $X$ under the
proportional-excess-loss reinsurance. Moreover, suppose that $I_{P}(X)$ ($I_{E}(X)$)
is the contribution of reinsurance against random claim
$X$ under the proportional ({\it or} the excess of loss)
reinsurance strategies. Then:
\begin{eqnarray}
P(X - {I_N}(X)\mathop  {\le}X - {I_{P}}(X))=P(X - {I_N}(X)\mathop
{\le}X - {I_{E}}(X)))=1.
\end{eqnarray}
\end{theorem}
\textbf{Proof.} To achieve the desired proof, it suffices to show
that  $P(A_1)=1$ ($P(A_2)=1$), where $A_1:=\{{I_N}(X)\mathop {\ge}
{I_{E}}(X)\}$ ($A_2:=\{{I_N}(X)\mathop {\ge} {I_{P}}(X)\}$). Now
consider the following two cases:
\begin{description}
     \item[(i)] Under excess of loss reinsurance, ${I_{E}}(X) =
     X-min(X,M);$ therefore:
\begin{eqnarray*}
  P(A_1) &=& P(A_1,X \le M) + P(A_1,X > M)\\
       &=& P(0 < X \le M) + P(X> M) = 1;
\end{eqnarray*}
     \item[(ii)] Under proportional reinsurance, ${I_{P}}(X) = (1 - \alpha
     )X;$ therefore:
\begin{eqnarray*}
  P(A_2) &=& P(A_2,X \le M)+ P(A_2,X > M)\\
       &=&P( X \le M) + P(X > M) =1.~\square
\end{eqnarray*}
\end{description}
From Theorem \eqref{compare-Stop-Proportiona-new} and properties
of VaR and the TVaR it can be concluded that the VaR and the TVaR
of the insurer contribution under proportional-excess-loss
reinsurance is less than for excess of loss and proportional
reinsurance strategies, i.e.,
 $\hbox{VaR}[X - {I_N}(X);p] \le \hbox{VaR}[X -
{I_{E}}(X);p]$ $ (\hbox{VaR}[X - {I_N}(X);p] \le \hbox{VaR}[X -
{I_{P}}(X);p])$ for all $p \in (0,1)$. It can be concluded that
$\hbox{TVaR}[X - {I_N}(X);p] \le \hbox{TVaR}[X -{I_{E}}(X);p]$ ($\hbox{TVaR}[X - {I_N}(X);p] \le \hbox{TVaR}[X -{I_{P}}(X);p],$)
for all $p \in (0,1)$.
\section{Estimating proportional-excess-loss reinsurance parameters}
This section considers proportional excess of loss reinsurance as
defined in Equation \eqref{Model-New-reinsurance}. An optimal
reinsurance strategy was derived by estimating unknown parameters
$\alpha$ and $M.$ First, the parameters were estimated by
maximizing the expected wealth for the insurer (reinsurer) using
an exponential utility function. Next, the estimated parameters
from the insurer and reinsurer were used as target estimators. A
Bayesian estimator was developed for the doubly-balanced loss
function for each parameter to maximize the expected exponential
utility of terminal wealth for the insurer and reinsurer.
Parameters  $\alpha$ and $M$ were first estimated using
exponential utility function to maximize the expected exponential
utility of the reinsurer's terminal wealth. Represent the surplus
of the insurer in the proportional excess of loss reinsurance
strategy as:
\begin{eqnarray}
\label{insurer-surplus}
\nonumber  U_t &=& u_0 + (1+\theta_0) E(\sum\limits_{i = 1}^{N(t)} {{Y_i}}) - \sum\limits_{i = 1}^{N(t)} {{Y_i}}\\
  &=& u_0 + \pi_0(t) - S(t),
\end{eqnarray}
where $u_0$ is the initial wealth of the insurer, random
variable $Y_i$ is the insurer portion of random
claim $X_i,$ $\theta_0$ is the safety factor, and $N(t)$ is the
Poisson process with intensity $\lambda.$ The expected wealth of
the insurer under the exponential utility
$u(x)=-e^{-\beta_0 x}$ is:
\begin{eqnarray}
\label{expecte-wealth-utility} E( { - \exp ( { - \beta_0 ( {{U_0}
+ \pi_0(t) - \sum\limits_{i = 1}^{N(t)} {{Y_i}} } )} )} ).
\end{eqnarray}
Using the definition for premium $\pi_0(t):$
\begin{eqnarray}
\pi_0(t) = (1 + \theta_0 ) \lambda t \left[ {\alpha \int_0^{{M}}
{xdF(x)}  + \alpha M\left[ {1 - F(M)} \right]} \right],
\end{eqnarray}
where $f(\cdot)$ and $F(\cdot)$ are the density and distribution
functions of random claim $X_i,$ respectively. Theorem
\eqref{alpha-M-zero-estimate} provides two estimators for $\alpha$
and $M,$ $\hat{\alpha}_0$ and $\hat{M}_0,$ that maximize the
expected wealth of the insurer Formula
\eqref{expecte-wealth-utility}.
\begin{theorem}
\label{alpha-M-zero-estimate} Suppose the surplus of the insurer
for proportional-excess-loss reinsurance strategy is calculated
using Equation \eqref{insurer-surplus}. Then, $\hat{\alpha}_0$ and
$\hat{M}_0$ maximize the expected exponential utility of the
insurer's terminal wealth from Equation
\eqref{expecte-wealth-utility} as:
\begin{eqnarray*}
  0 &=&-{\hat{\alpha}_0 \beta_0 }\hat{M}_0+\ln (1 + \theta_0 ), \\
  0 &=& - \beta_0 (1 + \theta_0 ) \lambda t\int_0^{{{\hat{M}}_0}} {xdF(x)}  - \beta_0 (1 + \theta_0 ) \lambda t\hat {M}_0\bar F(\hat{M}_0) \\
&& + \lambda \beta_0 t\int_0^{{\hat{M}_0}} {x {e^{\hat{\alpha}_0
\beta_0 x}}dF(x)}  + \lambda \beta_0 t \hat{M}_0{e^{\hat{\alpha}_0
\beta_0 \hat{M}_0}}\bar F(\hat{M}_0),
\end{eqnarray*}
where $\bar{F}(\cdot)$ is the survival function.
\end{theorem}
\textbf{Proof.}  Restate the expected exponential utility of the insurer's terminal wealth, (Equation \ref{expecte-wealth-utility}) as follows:
\begin{eqnarray*}
&&  - {e^{ - {\beta _0}({U_0} + {\pi _0}(t))}}E({e^{{\beta _0}\sum\limits_{i = 1}^{N(t)} {{Y_i}} }})\\
 &=&  - {e^{ - {\beta _0}({U_0} + {\pi _0}(t))}}{e^{\lambda t(E({e^{{\beta _0}Y}})-1)}}  \\
 &=&- e^{ ( { - \beta_0 ({U_0} + (1 + \theta_0
)\lambda t\left[ {\alpha \int_0^{{M}} {xdF(x)}  + \alpha M\bar
F(M)} \right]) }}\\
&&\times {e^{ \lambda t\left[ {\int_0^{{M}} {{e^{\alpha \beta_0
x}}dF(x)}  + {e^{\alpha \beta_0 M}}\bar F(M)} \right]} )}.
\end{eqnarray*}
Parameters $\alpha$ and $M$ maximize this expression and can be calculated by minimizing:
\begin{eqnarray}
\nonumber g_0(\alpha , M)&=&- \beta_0 (1 + \theta_0 )\lambda
t\left[ {\alpha \int_0^{{M}} {xdF(x)}  + \alpha M\bar
F(M)} \right]\\
&& + \lambda t\left[ {\int_0^{{M}} {{e^{\alpha \beta_0 x}}dF(x)} +
{e^{\alpha \beta_0 M}}\bar F(M)} \right].
\end{eqnarray}
Differentiating $g_0(\alpha , M)$ with respect to $\alpha$ and $M$
and setting them equal to zero produces:
\begin{eqnarray*}
\frac{{\partial g_0}}{{\partial \alpha }} &= & -  \beta_0 (1 + \theta ) \lambda t\int_0^{{M}} {xdF(x)}  - \beta_0 (1 + \theta ) \lambda tM\bar F(M) \\
&& +  \lambda \beta_0 t\int_0^{{M}} {x {e^{\alpha \beta_0
x}}dF(x)}  + \lambda \beta_0 t M{e^{\alpha \beta_0 M}}\bar
F(M)=0\\
\frac{{\partial g_0}}{{\partial M}} &=&  - \beta_0 (1 + \theta
)\alpha \lambda t\bar F(M) + \lambda \alpha \beta_0 t{e^{\alpha
\beta_0 M}}\bar F(M)=0.
\end{eqnarray*}
It is proven that the solutions to this for $\alpha$  and $M,$ $\hat{\alpha}_0$ and
$\hat{M}_0,$ minimize $g_0(\alpha, M).$ It must be shown that the following Hessian matrix at point
$(\hat{\alpha}_0 , \hat{M}_0)$ has a positive determinant and that the first argument ($a_{11}$) also  positive.
\begin{eqnarray}
\nonumber  \left(
{\begin{array}{*{20}{c}}
{\lambda t\int_0^{\hat{M}_0} {{\beta_0 ^2}{x^2}{e^{\hat{\alpha}_0 \beta_0 x}}dF(x)}  + \lambda t{\beta_0 ^2}{{\hat{M}_0}^2}{e^{\hat{\alpha}_0 \beta_0 M_0}}\bar F(\hat{M}_0)}&{\lambda {{\hat{\alpha}_0 }}{\beta_0 ^2}{t \hat{M}_0}{e^{\hat{\alpha}_0 \beta_0 \hat{M}_0}}\bar F(\hat{M}_0)}\\
{\lambda {{\hat{\alpha}_0 }}{\beta_0 ^2}{t \hat{M}_0}{e^{\hat{\alpha}_0 \beta_0 \hat{M}_0}}\bar F(\hat{M}_0)}&{\lambda {\hat{\alpha}_0}^2 {\beta_0 ^2} t {e^{\hat{\alpha}_0 \beta_0 \hat{M}_0}}\bar F(\hat{M}_0)}
\end{array}} \right).
\end{eqnarray}
This is arrived at using straightforward calculation.  $\square$

When $\hat{\alpha}_0>1(<0),$ it must be projected into $[0,1].$
Now estimate unknown parameters $\alpha$ and $M$ in the
proportional-excess-loss reinsurance strategy Equation
\eqref{Model-New-reinsurance} to maximize the expected exponential
utility function ($u(x)=-e^{-\beta_1 x}$) of the reinsurance
wealth. Suppose the surplus of reinsurer company under the
proportional-excess-loss reinsurance strategy is:
\begin{eqnarray}
\label{surplus-reinsurance}
U_t^* &=& u_0^* + \pi_1(t) -
\sum\limits_{i = 1}^{N(t)} {I({X_i})}
\end{eqnarray}
where $u^*_0$ is the initial wealth of the reinsurer,
random variable $I({X_i})$ represents the reinsurer portion
against random claim $X_i,$ $\pi_1(t)$ is premium of the reinsurance
strategy in time $t$, and $N(t)$ is a Poisson process with intensity
$\lambda.$ Under the expectation premium principle with safety
factor $\theta_1,$ premium $\pi_1(t)$ can be restated as:
\begin{eqnarray*}
(1 + \theta_1 ) \lambda t \left[ {(1-\alpha) \int_0^{M} {xdF(x)}
+\int_{{M}}^\infty  {(x - \alpha M)dF(x)} ]} \right],
\end{eqnarray*}
where $f(\cdot)$ is the density function of
random claim $X_i.$ The expectation of reinsurer wealth using exponential utility function $u(x)=-e^{-\beta_1 x}$ is:
\begin{eqnarray}
\label{reinsurance-target}
E(-exp( - {\beta _1}(u_0^*
+ \pi_1(t)-\sum\limits_{i = 1}^{{N}(t)} {I({X_i})}))).
\end{eqnarray}
Theorem \eqref{alpha-M-one-estimate} provides two estimators for
$\alpha$ and $M,$ $\hat{\alpha}_1$ and $\hat{M}_1,$ that maximize
Equation \eqref{reinsurance-target}.
\begin{theorem}
\label{alpha-M-one-estimate} Suppose the surplus for a reinsurance
company under the proportional-excess-loss reinsurance strategy
can be represented by Equation \eqref{surplus-reinsurance}. Then,
$\hat{\alpha}_1$ and $\hat{M}_1$ which maximize the expected
exponential utility of the reinsurer's terminal wealth given by
Equation \eqref{reinsurance-target}, can be found as:
\begin{eqnarray*}
0 =&-& \int_0^{{\hat{M}_1}} {{\beta _1}x{e^{  {\beta _1}(1 - \hat{\alpha}_1 )x}}dF(x)}  - \int_{{\hat{M}_1}}^\infty  {{\beta _1}\hat{M}_1{e^{ {\beta _1}(x - \hat{\alpha}_1 \hat{M}_1)}}dF(x)}\\
 &+& {\beta _1}(1 + \theta_1 )\int_0^{{\hat{M}_1}} {xdF(x)}  + {\beta _1}(1 + \theta_1 )\int_{{\hat{M}_1}}^\infty  { \hat{M}_1dF(x)}\\
0=&-& \int_{{\hat{M}_1}}^\infty  {{\beta _1}\hat{\alpha}_1 {e^{ -
{\beta _1}(x - \hat{\alpha}_1 \hat{M}_1)}}dF(x)}  + {\beta _1}(1 +
\theta )\hat{\alpha}_1 (1 - F(\hat{M}_1))
\end{eqnarray*}
\end{theorem}
\textbf{Proof.} Parameters $\alpha$ and $M$ maximize the expected
exponential utility of the reinsurer's terminal wealth in Equation
\eqref{reinsurance-target} and can be found by minimizing the
following expression:
\begin{eqnarray*}
 g_1(\alpha , M) = &&\int_0^{{M}} {{e^{ {\beta _1}(1 - \alpha)x}}dF(x)}  \\
&+& \int_{{M}}^\infty  {{e^{  {\beta _1}(x - \alpha M)}}dF(x)} \\
& - & {\beta _1}(1 + \theta_1 )\int_0^{{M}} {(1 - \alpha )xdF(x)}\\
&-& {\beta _1}(1 + \theta_1 )\int_{{M}}^\infty  {(x - \alpha
M)dF(x)}
\end{eqnarray*}
Differentiating $g_1(\alpha , M)$ with respect to $\alpha$ and $M$
and setting them equal to zero produces:
\begin{eqnarray*}
\frac{{\partial g_1}}{{\partial \alpha}} =&-& \int_0^{{M}} {{\beta _1}x{e^{  {\beta _1}(1 - \alpha)x}}dF(x)}  \\
&-& \int_{{M}}^\infty  {{\beta _1}M{e^{ {\beta _1}(x - \alpha M)}}dF(x)}\\
 &+& {\beta _1}(1 + \theta_1 )\int_0^{{M}} {xdF(x)} \\
  &+& {\beta _1}(1 + \theta_1 )\int_{{M}}^\infty  { MdF(x)}=0\\
\frac{{\partial g_1}}{{\partial M}} =&-& \int_{{M}}^\infty {{\beta
_1}\alpha {e^{ - {\beta _1}(x - \alpha M)}}dF(x)} \\
&+& {\beta _1}(1
+ \theta )\alpha (1 - F(M))=0.
\end{eqnarray*}
The proof shows that the solutions of this equation for $\alpha$  and $M,$ $\hat{\alpha}_1$ and
$\hat{M}_1,$ minimize $g_1(\alpha, M).$ It must be shown that the following Hessian matrix at point
$(\hat{\alpha}_1, \hat{M}_1)$ has a positive determinant and $a_{11}>0$:
\begin{eqnarray*}
H_1(\hat{\alpha}_1, \hat{M}_1) &=& \left( {\begin{array}{*{20}{c}}
{{a_{11}}}&{{a_{12}}}\\
{{a_{21}}}&{{a_{22}}}
\end{array}} \right),
\end{eqnarray*}
where
\begin{eqnarray*}
{a_{11}} &=& \int_0^{{\hat{M}_1}} {\beta _1^2{x^2}{e^{  {\beta _1}(1 - \hat{\alpha}_1 )x}}dF(x)} \\
 &&+ \int_{{\hat{M}_1}}^\infty  {\beta _1^2{\hat{M}_1^2}{e^{  {\beta _1}(x - \hat{\alpha}_1 \hat{M}_1)}}dF(x)} \\
{a_{12}} &=& {a_{21}} = (-1 + {\beta _1}\hat{\alpha}_1 \hat{M}_1)\int_{{\hat{M}_1}}^\infty  {{\beta _1}{e^{  {\beta _1}(x - \hat{\alpha}_1 \hat{M}_1)}}dF(x)} \\
&& + {\beta _1}(1 + \theta )(1 - F(\hat{M}_1))\\
{a_{22}} &=& \int_{{\hat{M}_1}}^\infty  {\beta
_1^2{\hat{\alpha}_1^2}{e^{ {\beta _1}(x -\hat{\alpha}_1
\hat{M}_1)}}dF(x)}\\
&& + {\beta _1}\hat{\alpha}_1 {e^{ {\beta _1}(1 -
\hat{\alpha}_1 )\hat{M}_1}}f(\hat{M}_1) - {\beta
_1}\hat{\alpha}_1(1 + \theta )f(\hat{M}_1).
\end{eqnarray*}
the positivity of the determinant of the Hessian matrix cannot
be established and must be verified in practice; however, it is evident that $a_{11}>0.$
$\square$

Thus far, the optimal reinsurance strategy has been defined for
the insurer and reinsurer to integrate the results and define an
optimal reinsurance strategy that considers the interests of both
parties. A Bayesian estimator was developed for $\alpha$ and $M$
for the doubly-balanced loss function in Equation
\eqref{doubly-balance-loss}. Estimators $\hat{\alpha}_0$ and
$\hat{\alpha}_1$ ($\hat{M}_0$ and $\hat{M}_1$) are target
estimators for the doubly-balanced loss function.

For convenience, $Z_i=I(X_i)$ . Lemma \eqref{joint-density}
provides a cumulative distribution function and density function
for conditional random variable $Z|(\theta ,\alpha ,M).$
\begin{lemma}
\label{joint-density} Suppose $X| \theta $ has continuous
distribution function $F_{X| \theta }(\cdot)$ and continuous
density function $f_{X|\theta}(\cdot).$ Moreover, suppose that
$Z_1,\cdots,Z_n|(\theta, \alpha, M)$ are a sequence of i.i.d.
random variables with common density function ${f_{Z|(\theta,
\alpha, M)}}(\cdot).$ Then, the joint density function of
$Z_1,\cdots,Z_n|(\theta, \alpha, M)$ can be represented as:
\begin{eqnarray*}
f({z_1},\cdots,{z_n}\left| \theta, \alpha, M  \right.) = &&{\left(
{\frac{1}{{1 - \alpha }}} \right)^{{n_1}}}\prod\limits_{i =
1}^{{n_1}} {{f_{X\left| \theta  \right.}}(\frac{{{z_i}}}{{1 -
\alpha }})}\\
&&\times \prod\limits_{i = {n_1} + 1}^n {{f_{X\left| \theta
\right.}}({z_i} + \alpha M)},
\end{eqnarray*}
where $n_1$ is the number of $z_i$s that is less than or equal to $(1-\alpha)M$.
\end{lemma}
\textbf{Proof.} For one sample
$Z\left| (\theta, \alpha, M) \right.$ the distribution function is:
\begin{eqnarray*}
{F_{Z\left| {\theta ,\alpha ,M} \right.}}(z) &=& P(Z \le z)\\
 &=& P((1 - \alpha )X \le z,X \le M)\\
 && + P(X - \alpha M \le z,X > M)\\
&=& P(X \le \min \{ \frac{z}{{1 - \alpha }},M\} )\\
&& + P(M < X \le z + \alpha M)\\
&=& {F_X}(\min \{ \frac{z}{{1 - \alpha }},M\} )\\
&& + {F_X}(z + \alpha M) - {F_X}(M)\\
&=& {F_X}(\frac{z}{{1 - \alpha }})I_{(-\infty,~(1 - \alpha
)M]}(z)\\
&&+{F_X}(z + \alpha M)I_{((1 - \alpha )M,~\infty)}(z),
\end{eqnarray*}
where $I_A(x)$ stands for the indicator function. Differentiating
$F(z)$ with respect to $z$ leads to:
\begin{eqnarray*}
 {f_{Z\left| {\theta ,\alpha ,M} \right.}}(z)
=&&\frac{1}{{1 - \alpha }}{f_X}(\frac{z}{{1 - \alpha
}})I_{(-\infty,~(1 - \alpha )M]}(z)\\
&&+{f_X}(z + \alpha M)I_{((1 -
\alpha )M,~\infty)}(z).
\end{eqnarray*}
Suppose that $n_1$ ($0\leq n_1\leq n$) represents the number
of $z_i$s that less than or equal to $(1-\alpha)M.$ Joint density function for an
independent sequence of random variables obtained by multiplying
their marginal density functions is the desired proof. $\square$

Lemma \eqref{posterior-density} develops the joint posterior
distribution for $(\theta,\alpha, M)$ given random sample
${{Z_1},\cdots,{Z_n}}.$
\begin{lemma}
\label{posterior-density} Suppose $Z_1,\cdots,Z_n|(\theta,\alpha,
M)$  are a sequence of i.i.d. random variables with common density
function ${f_{Z\left| {\theta ,\alpha ,M} \right.}}(z).$ Moreover,
suppose that $\pi_1 {(\Theta)}$, $\pi_2 {(\rm \mathcal{A})},$ and
$\pi_3 {({\rm \mathcal{M}})}$ are prior distributions for
$\theta$, $\alpha,$ and $M,$ respectively. Then, the joint
posterior distribution for $\left( {\theta ,\alpha ,M\left|
{{Z_1},\cdots,{Z_n}} \right.} \right)$ is:
\begin{eqnarray*}
\frac{{{{\left( {\frac{1}{{1 - \alpha }}}
\right)}^{{n_1}}}\prod\limits_{i = 1}^{{n_1}} {{f_{X\left| \theta
\right.}}\left( {\frac{{{z_i}}}{{1 - \alpha }}} \right)}
\prod\limits_{i = {n_1} + 1}^n {{f_{X\left| \theta
\right.}}\left( {{z_i} + \alpha M} \right)} }{\pi_1
{(\theta)}}{\pi_2 {(\alpha)}}{\pi_3 {(M)}}}{{\int\limits_{\rm M}
{\int\limits_{\rm A} {\int\limits_\Theta  {{{\left( {\frac{1}{{1 -
\alpha }}} \right)}^{{n_1}}}\prod\limits_{i = 1}^{{n_1}}
{{f_{X\left| \theta  \right.}}\left( {\frac{{{z_i}}}{{1 - \alpha
}}} \right)} \prod\limits_{i = {n_1} + 1}^n {{f_{X\left| \theta
\right.}}\left( {{z_i} + \alpha M} \right)}{\pi_1
{(\theta)}}{\pi_2 {(\alpha)}}{\pi_3 {(M)}}d\theta d\alpha dM} } }
}}
\end{eqnarray*}
where $n_1$ is the number of $z_i$s that less than or equal to
$(1-\alpha)M$.
\end{lemma}
\textbf{Proof.} The joint density function of ${{Z_1},\cdots,{Z_n}}|(\theta ,\alpha ,M)$
plus the prior distributions for $\theta$, $\alpha,$ and $M$ are the desired proof.
$\square$

The marginal density functions for $\left( {\alpha \left|
{{Z_1},\cdots,{Z_n}} \right.} \right)$ and $\left( {M \left|
{{Z_1},\cdots,{Z_n}} \right.} \right)$ are
\begin{eqnarray*}
  \pi (\alpha \left| {{Z_1},\cdots,{Z_n}} \right.) &=& \int \limits_{\Theta} \int\limits_{\rm \mathcal{M}} {\pi (\theta, \alpha ,M\left| {{Z_1},\cdots,{Z_n}} \right.)\,dM d\theta}; \\
  \pi (M\left| {{Z_1},\cdots,{Z_n}} \right.) &=& \int \limits_{\Theta}\int\limits_{\rm \mathcal{A}} {\pi (\theta, \alpha ,M\left| {{Z_1},\cdots,{Z_n}} \right.)\,d\alpha d\theta }.
\end{eqnarray*}
Theorem \eqref{Bayes-Estimators} provides the Bayesian estimator
for $\alpha$ and $M$ for the doubly-balanced loss function in
Equation \eqref{doubly-balance-loss}.
\begin{theorem}
\label{Bayes-Estimators} Suppose $Z_1,\cdots,Z_n|(\theta, \alpha,
M)$ are a sequence of i.i.d. random variables with common density
function ${f_{Z\left|(\theta, \alpha, M)\right.}}(z)$. Moreover,
suppose that $\pi_1 {(\Theta)}$, $\pi_2 {(\rm \mathcal{A})},$ and
$\pi_3 {({\rm \mathcal{M}})}$ are prior distributions for
$\theta$, $\alpha,$ and $M,$ respectively. Then, the Bayesian
estimators for $\alpha$ and $M$ for the square error
doubly-balanced loss function, prior distribution $\pi,$ and
target estimators $\hat{\alpha}_0,~\hat{\alpha}_1$ and
$\hat{M}_0,~\hat{M}_1$ are
\begin{eqnarray*}
{\hat{\alpha} _{\pi ,{\omega _1},{\omega _2}}}&=& {\omega
_1}{\hat{\alpha} _0} + {\omega _2}{\hat{\alpha} _1} + (1 - {\omega
_1} - {\omega _2}){E_\pi
}(\mathcal{A} \left| z \right.),\\
 {\hat{M} _{\pi ,{\omega
_1},{\omega _2}}}&=& {\omega _1}{\hat{M} _0} + {\omega _2}{\hat{M}
_1} + (1 - {\omega _1} - {\omega _2}){E_\pi }(\mathcal{M} \left| z
\right.).
\end{eqnarray*}
\end{theorem}
\textbf{Proof.} The results of Lemma \eqref{joint-density}, Lemma
\eqref{posterior-density}, Theorem \eqref{double balance} and
Corollary \eqref{square loss} provide the desired proof.
 $\square$
\section{Simulation study}
This section provides two numerical examples to show how the above
findings can be applied in practice. It develops {\bf {(i)}}
estimators for $\alpha$ and $M,$ $\hat{\alpha}_0$ and $\hat{M}_0,$
so that insurer wealth is maximized; {\bf {(ii)}} estimators for
$\alpha$ and $M,$ $\hat{\alpha}_1$ and $\hat{M}_1,$ so that
reinsurer's wealth is maximized; {\bf {(iii)}} Bayesian
estimators for $\alpha$ and $M$ for the square error
doubly-balanced loss function for prior distributions $\alpha\sim
Beta(2,2)$ and $M\sim Exp(2),$ and target estimators
$\hat{\alpha}_0,~\hat{\alpha}_1$ and $\hat{M}_0,~\hat{M}_1.$
\begin{example}
Suppose 4.117, 1.434, 0.453, 3.333, 0.456, 0.0637, 0.145, 0.211,
3.618, 5.467 is a random sample generated from an exponential
distribution with intensity $1.$ Moreover, suppose that
$Beta(2,2)$ and $Exp(2)$ are prior distribution functions for
parameters $\alpha$ and $M,$ respectively.
\end{example}
The following provides practical steps to find the optimal
proportional-excess-loss reinsurance strategy.
\begin{description}
    \item[Step 1:] Assuming $\beta_0 = 2$ and $\theta_0 = 0.8,$ in
    Theorem \eqref{alpha-M-zero-estimate}, lead to $\hat{\alpha}_0 = 0.27$ and $\hat{M}_0 = 1.08;$
    \item[Step 2:] Assuming $\beta_1=0.2$ and $\theta_1=0.3,$ in
    Theorem \eqref{alpha-M-one-estimate}, lead to $\hat{\alpha}_1=0.38$ and
    $\hat{M}_1=37.001$ where $\det(H_1(0.38,37.001))>0;$
    \item[Step 3:] Suppose $0.453, 0.456, 0.0637, 0.145,
    0.211$ in the random sample are $\le (1-\alpha)M.$ Moreover, suppose that $Beta(2,2)$ and $Exp(2)$ are prior distribution functions for $\alpha$ and $M,$ respectively. Application of
Corollary \eqref{square loss} leads to the Bayesian estimators for
$\alpha$ and $M:$
\begin{eqnarray*}
{\hat{\alpha} _{\pi ,{\omega _1},{\omega _2}}} &=&0.27 {\omega _1}  + 0.38{\omega _2} + 0.6(1 - {\omega _1} - {\omega _2})\\
&=&0.6-0.33\omega_1-0.22\omega_2;\\
 {\hat{M}_{\pi ,{\omega _1},{\omega _2}}} &=& 1.08{\omega _1}
+ 37.001{\omega _2}+ 0.78(1 - {\omega _1} -{\omega _2})\\
&=&0.78+0.3\omega_1+36.221\omega_2,
\end{eqnarray*}
where, under boundary conditions $\omega_1$ and $\omega_2$
(i.e., $\omega_1~\&~\omega_2\in[0,1]$ and
$\omega_1+\omega_2\leq1$), both estimators are positive.
\end{description}
Table 1 shows Bayesian estimators ${\hat{\alpha} _{\pi ,{\omega
_1},{\omega _2}}}$ and ${\hat{M}_{\pi ,{\omega _1},{\omega _2}}}$
for different values of $\omega_1$ and $\omega_2$ that
satisfy the boundary conditions for $\omega_1$ and $\omega_2.$
\begin{center}
\scriptsize Table 1: Bayes estimators ${\hat{\alpha} _{\pi
,{\omega _1},{\omega _2}}}$ and ${\hat{M}_{\pi ,{\omega
_1},{\omega _2}}}$ for some different values of $\omega_1$ and $\omega_2.$\\
\begin{tabular}{c c c c c}
\hline
  & &  & ~~~~~~~~~~~~~~~~~~~~~~~~~~~~Bayes estimator\\ \cline{4-5}
${\omega_1}$ & ${\omega_2}$ & $1-\omega_1-\omega_2$ &
${\hat{\alpha} _{\pi ,{\omega _1},{\omega _2}}}$ & ${\hat{M}_{\pi
,{\omega _1},{\omega _2}}}$  \\\hline
0.1&   0.1&    0.8 &0.545  &  4.43      \\
0.1&    0.2&    0.7&0.523    &  8.05      \\
0.1&    0.3&    0.6&0.501    & 11.67     \\
0.1&    0.4&    0.5&0.479    & 15.29     \\
0.1&    0.5&    0.4&0.457    & 18.92    \\
0.1&    0.6&    0.3&0.435    & 22.54     \\
0.1&    0.7&    0.2&0.413    & 26.16      \\
0.1&    0.8&    0.1&0.391    & 29.78     \\
0.1&    0.9&    0&0.369  & 33.40      \\   \hline
0.1&    0.1&    0.8&0.545    & 4.432     \\
0.2&    0.1&    0.7&0.512    & 4.462     \\
0.3&    0.1&    0.6&0.479    & 4.492      \\
0.4&    0.1&    0.5&0.446    & 4.522     \\
 0.5&   0.1&    0.4&0.413    & 4.552     \\
0.6&    0.1&    0.3&0.380    & 4.582     \\
0.7&    0.1&    0.2&0.347    & 4.612     \\
0.8&    0.1&    0.1&0.314    & 4.642     \\
0.9&    0.1&    0&0.281  & 4.672     \\
\hline
\end{tabular}
\end{center}
As one may observe from result of Table 1, choice of
$\omega_1,~\omega_2$ have a big impact on estimated $M$ and do not
have such impact on estimated $\alpha.$

Using result of Table 1, one may determine the desired optimal
proportional-excess-loss reinsurance strategy.

The following example assume $\omega_1=0.25$, $\omega_2=0.15$ in
Theorem \eqref{Bayes-Estimators} provides the optimal
proportional-excess-loss reinsurance strategy for some different
claim size distributions.

\begin{example}
Suppose $X_1,\cdots,X_{100}$ is a sequence of random sample from
distributions given by the first column of Table 2. Moreover,
suppose that, for each claim size distribution, prior
distributions for $\alpha$ and $M$ are given by Table 2.
\end{example}
For each claim size distribution, we generate random sample
$X_1,\cdots,X_{100},$ 100 times and estimate parameters $\alpha$
and $M,$ for each iteration. Table 2 represents mean (and standard
deviation) of Bayes estimator of $\alpha$ and $M$ for such 100
iterations.

To estimate unknown parameters $\alpha$ and $M$ using the Bayesian
method, we need initials to categorized data into two groups and
prior distribution functions for $\alpha$ and $M$. Prior
distribution functions given by the second and third columns of
Table 2.
\begin{center}
\scriptsize{Table 2: Random claim and prior distributions
accompanied with posterior's mean and standard deviation.
\begin{tabular}[c]{c c c c c c}
\hline Claim size & prior &prior& The mean (SD)&The mean (SD)\\
 distribution& for $\alpha$ & for $M$  & for $\alpha$ & for $M$ \\
 \hline
EXP(1)          &Beta(2,2)      &EXP(2)     &  0.5189   &3.6915  \\
&&&(0.03300)&(0.04020)\\
EXP(4)          &Beat(2,2)      &EXP(2)     &  0.4306   &0.7456  \\
&&&(0.03960)&(0.00002)\\
EXP(8)          &Beta(3,2)      &Gamma(2,2) &  0.4402   &1.0748 \\
&&&(0.03660)& (0.00001)\\
Weibull(2,1)    &Beta(2,4)      &Gamma(3,2) &  0.5458   &1.3813 \\
&&&(0.01500)& (0.00003)\\
Weibull(4,1)    &Beat(5,2)      &Gamma(2,4) &  0.5464   &0.9142 \\
&&&(0.44160)& (0.02340)\\
Weibull(2,4)    &Uniform(0,1)   &Gamma(3,4) &  0.7612   &2.4772 \\
&&&(0.00780)& (0.02340)\\
\hline
\end{tabular}}\\
\normalsize
\end{center}
The small standard deviation of these estimators shows that the
estimation method is an appropriate method to use with the
different samples. Moreover, as one may observe from result of
Table 2, choice of prior distributions for $\alpha$ and $\beta$
have a sufficient impact on estimated $M$ and do not have such
impact on estimated $\alpha.$

The mean represented in Table 2 can be considered as a Bayesian
estimator for $\alpha$ and $M$ and determine optimal
proportional-excess-loss reinsurance.
\section{Conclusion}
This study combined excess of loss and proportional reinsurance
strategies to introduce a new reinsurance strategy, say
proportional-excess-loss reinsurance. This optimal reinsurance
strategy has been achieved by estimating unknown parameters for
the proportional-excess-loss reinsurance strategy such that the
expected exponential utility of the insurer's and reinsurer's
terminal wealth are maximized, simultaneously.

The new proportional-excess-loss reinsurance strategy can be
extended situations where the reinsurance strategy has more than
two unknown parameters. Then, unknown parameters have been
estimated from more optimal criteria.

\end{document}